\documentclass[twocolumn,prl,showpacs,superscriptaddress,byrevtex]{revtex4-1}
\usepackage{epsf,graphicx,amssymb}
\usepackage[usenames]{color}
\usepackage{natbib}
\usepackage{float}
\usepackage{amsmath,amssymb}
\usepackage{color}

\begin{document}

\title{Direct numerical simulations of capillary wave turbulence}
\author{Luc Deike}
\affiliation{Univ Paris Diderot, Sorbonne Paris Cit\'e, MSC, UMR 7057 CNRS, F-75 013 Paris, France, EU}
\affiliation{Scripps Institution of Oceanography, University of California San Diego, La Jolla, California}
\author{Daniel Fuster}
\affiliation{Institut Jean le Rond d'Alembert, Universit\'e Pierre et Marie Curie, UMR 7190, Paris, France, EU}
\author{Michael Berhanu}
\affiliation{Univ Paris Diderot, Sorbonne Paris Cit\'e, MSC, UMR 7057 CNRS, F-75 013 Paris, France, EU}
\author{Eric Falcon}
\affiliation{Univ Paris Diderot, Sorbonne Paris Cit\'e, MSC, UMR 7057 CNRS, F-75 013 Paris, France, EU}

\date{\today}
\begin{abstract}  
This work presents Direct Numerical Simulations of capillary wave turbulence solving the full 3D Navier Stokes equations of a two-phase flow. When the interface is locally forced at large scales, a statistical stationary state appears after few forcing periods. Smaller wave scales are generated by nonlinear interactions, and the wave height spectrum is found to obey a power law in both wave number and frequency in good agreement with weak turbulence theory. By estimating the mean energy flux from the dissipated power, the Kolmogorov-Zakharov constant is evaluated and found to be compatible with the exact theoretical value. The time scale separation between linear, nonlinear interaction and dissipative times is also observed. These numerical results confirm the validity of weak turbulence approach to quantify out-of equilibrium wave statistics.
\end{abstract}


\maketitle

Wave turbulence aims to provide a general description of a set of weakly nonlinear interacting waves. The theoretical framework of weak-wave turbulence has been widely applied to very different physical situations such as gravity and capillary waves at the surface of a liquid, internal waves in the ocean and the atmosphere, flexural waves on a plate, optical waves and magneto-hydrodynamical waves \cite{ZakharovBook,NazarenkoBook,NewellReview}.  Experimental and numerical results in various systems show that the description provided by weak turbulence has limitations. For instance, the existence of dissipation at all scales has an influence on the energy spectrum measured in capillary wave turbulence \cite{DeikePre2013} as well as in flexural wave turbulence on plates \cite{Humbert2013,Miquel2013}. In the case of ocean gravity waves, wave breaking is known to be the main dissipation source appearing at various scales \cite{MEL02Nat,MelvilleRomero,MelvilleSutherland}. The existence and influence of coherent 
structures among a set of random waves is also an open question in wave turbulence \cite{Zakharov2007,ResidoriReview,Miquel2013Prl}. Direct Numerical Simulations (DNS) are an appealing tool to quantify the influence of different processes on wave turbulence and test the various theoretical hypotheses separately.

Capillary wave is one of the simplest systems to study wave turbulence. However, this regime often interacts with gravity wave turbulence in experiments and these mutual interactions remain an open question \cite{NewellZakharov2008,DeikePre2013}. The numerical investigation of purely capillary wave turbulence finds application probing the validity range of weak turbulence theory in experiments, and to improve our understanding of the influence of capillary waves at the ocean surface regarding dissipation, air-water exchanges \cite{MelvilleSutherland,FM98,CoxMunk1954b}, or microwave remote sensing techniques of the ocean surface  \cite{CoxMunk1954b,Phillips1988}.

The main result of capillary wave turbulence is the existence of a direct energy cascade. In terms of the spatial wave power spectrum $S_{\eta}(k)$, where $k$ is the wave number, the Kolmogorov-Zakharov spectrum reads  \cite{Zakharov67} 
\begin{equation}
S_{\eta}(k)=C_k^{KZ}\epsilon^{1/2}\left(\gamma/\rho\right)^{-3/4}k^{-15/4},
\label{eqKZ}
\end{equation}
with  $\epsilon$ the mean energy flux, $\gamma$ the surface tension and $\rho$ the liquid density. $C_k^{KZ}$ is the non-dimensional Kolmogorov-Zakharov (KZ) constant that can be explicitly calculated \cite{Zakharov67,Pushkarev2000}. This direct energy cascade has been widely explored experimentally finding good agreement with theory for the frequency (or wave-number) scaling of the wave spectrum \cite{Holt1996,Putterman1996,Brazhnikov02,Falcon07a,Falcon0g,Xia10,Berhanu2013}.  Recent works address the influence of dissipation on wave turbulence \cite{DeikePre2013} and the coexistence of anisotropic structures and wave turbulence \cite{Berhanu2013}. Numerical simulations is a powerful tool to answer to these questions, notably it allows to access quantities difficult to measure experimentally. In comparison to the wide literature available for experimental measurements, there are only a few numerical studies on capillary wave turbulence. These studies can be typically classified in two different groups: kinetic 
equation simulations \cite{Resio,ZakharovBook,Kolmakov06} and weakly nonlinear hamiltonian dynamics simulations \cite{Pushkarev1996,Pushkarev2000}. Both approaches remain limited to weakly nonlinear situations and cannot investigate the possible influence of air and water flow on the waves. The reason why more complete models have not been tested is that solving the full Navier-Stokes equations in multiphase flow is a numerical challenge. Only thanks to the recent development of numerical methods \cite{POP07} it is now possible to perform long wave turbulence simulations, in order to obtain representative statistics, with relative high resolution in terms of wave number.

In this letter, we present the first observation of the direct energy cascade in capillary wave turbulence from the numerical solution of the full 3D Navier-Stokes equations. The numerical method and physical configuration are first introduced and the stationary state is characterized. We show the ability of the simulations to capture the propagation of capillary waves. The obtained space-time wave spectrum is compared with the classical dispersion relation. The wave spectrum both in wave number or frequency is found to obey a power law in good agreement with the weak turbulence theory from which the KZ constant can be estimated. The nonlinear interaction time scale $\tau_{nl}$ is estimated and is found to scale as $\tau_{nl}\sim k^{-3/4}$, in agreement with wave turbulence theory. Finally, we show that the wave turbulence inertial range is defined by $\tau_l\ll\tau_{nl}\ll\tau_d$, where $\tau_l$ is the linear propagation time and $\tau_d$ the dissipation time.

\begin{figure}
\includegraphics[scale=0.42]{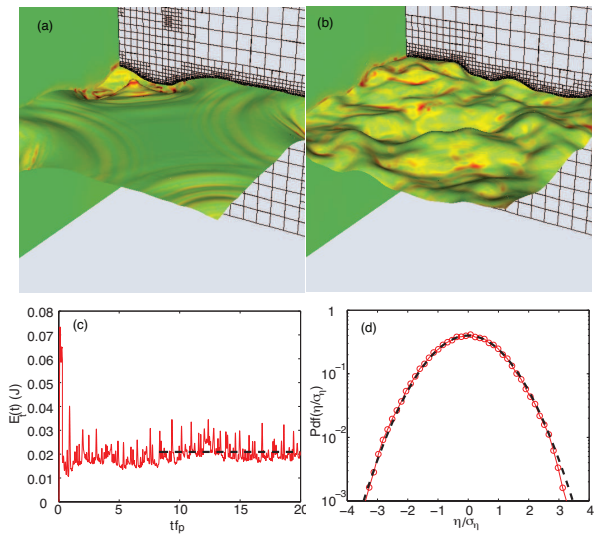}
\quad\caption[]{(a,b) Snapshot of the wave interface $\eta(x,y)$ at time $tf_p=0.3$ and $tf_p=9.1$. (c) Total wave energy $E_t$ as a function time $tf_p$. A stationary state is reached for $tf_p>5$ and the mean value is indicated by dashed line. (d) PDF of the wave height $\eta/\sigma_{\eta}$. Dashed line is the normalized gaussian.}
\label{Stat_im}
\end{figure}

We solve the 3D two-phase Navier Stokes equations accounting for surface tension and viscous effects using the open source solver Gerris \cite{POP03,POP07}. This solver has been successfully used in multiphase problems like atomization, the growth of instabilities at the interface \cite{Fuster2013}, wave breaking \cite{Fuster} or splashing \cite{PopinetVK}. The physical properties of the two phases are those of air and water. Gravity is not present. The simulation domain is a cube of length $L=1$ m with periodic boundary conditions in the $x$ and $y$ horizontal directions. In the vertical direction $z$ slipped wall (symmetry) conditions are imposed. The interface between the liquid and gas phase is initially placed at the half plane $z=0$ (the water depth is $h=0.5L$). It is perturbed locally introducing a source in the momentum equation. This source is obtained from the linear wave solution \cite{Lamb} corresponding to a forcing on the interface elevation $\eta$: $\eta(x,y,t)=\alpha(x,y)\sum_{
i=0}^{4}\eta_0\cos{(\vec{k_i}\cdot\vec{x}-\omega_i t)}$; and on the liquid velocity $\vec{v}$ ($\vec{v}=\vec{\nabla}\phi$, $\phi$ the velocity potential): $\phi(x,y,z,t)=\alpha(x,y)\sum_{i=0}^{4}\left(-\eta_0\frac{\omega_i}{k_i}\frac{\cosh k_i(z+h)}{\cosh(k_i h)}\right)\cos{(\vec{k_i}\cdot\vec{x}-\omega_i t)}$, where $\eta_0$ is the wave amplitude. The forcing modes are $k_0=k_p$, $k_1=1.4k_p$, $k_2=1.2k_p$, $k_3=0.8k_p$, $k_4=0.6k_p$, with $k_p=2\pi/(0.4L)$ the central forcing wave number and $\omega_i=\sqrt{(\gamma/\rho)k_i^3 \tanh(kh)}$ the corresponding frequency given by the linear dispersion relation of capillary waves. The forcing is located in space through the gaussian function $\alpha(x,y)=\exp{(-(x-x_c)^2-(y-y_c)^2)/2r^2}$, with $r=0.15L$, $x_c=y_c=0.25L$. Note that the forcing area size has no influence on the generated wavelength. We expect capillary waves to propagate according to the linear dispersion relation $\omega^2=(\gamma/\rho)k^3$ \cite{Lamb}. The maximal resolution in the simulations presented here corresponds to 256$\times$256 grid on the 
interface. Adaptive mesh refinement is used in order to decrease the resolution in the bulk and to reduce the computational time.   However, despite the use of adaptive mesh refinement techniques, the high computational cost of the method has impeded to refine the grid to a level where the numerical viscosity naturally introduced by the numerical schemes becomes negligible compared to the physical viscosity \cite{SMDeike}. Note that the effect of artificial numerical dissipation is also present in previous numerical computations presented in the literature with pseudo-spectral methods \cite{Pushkarev1996,Pushkarev2000}.

Figure \ref{Stat_im} (a,b) shows two snapshots of the interface $\eta(x,y)$ during the first period of forcing (Fig. \ref{Stat_im}a) and after few forcing periods (Fig. \ref{Stat_im}b). The forcing area is clearly visible on Fig. \ref{Stat_im}a (top left corner), where we see waves propagate from the circular region where the source is applied and from others corners due to the periodic boundary conditions. After a few forcing wave periods, (Fig. \ref{Stat_im}b) the wave field displays random features with a wide range of spatial scales. The wave field nonlinearity is estimated by the typical steepness $r=\sigma_{\eta}k_p\approx 0.3$, with $\sigma_{\eta}$ the rms value of the wave amplitude. Figure \ref{Stat_im}c shows the total wave energy $E_t=E_s+E_c$ as a function of dimensionless time $t f_p$. The kinetic and potential energy on the domain volume $\Omega$ are computed from the liquid velocity $v$, and the surface area, $A(x,y)$, as $E_c=\int \frac{1}{2}\rho v^2d\Omega$ and $E_s=\gamma A(x,y)$. 
After a short transient state, the wave energy reaches a stationary state ($tf_p>5$) where the wave energy fluctuates around a constant mean value (displayed by a dashed line in Fig. \ref{Stat_im}b). Wave statistics are obtained in the time interval $t = [10:20] f_p$, with $f_p=\omega_p/(2\pi)$. We now focus on the statistical and dynamical properties of the waves during this stationary regime. Figure \ref{Stat_im}d shows the probability density function (PDF) of the wave height $\eta/\sigma_{\eta}$ during this stationary stage. Gaussian statistics are observed, meaning that the nonlinear effects are weak enough to not induce a significant asymmetry on the capillary waves.

\begin{figure}
\begin{center}
\includegraphics[scale=0.42]{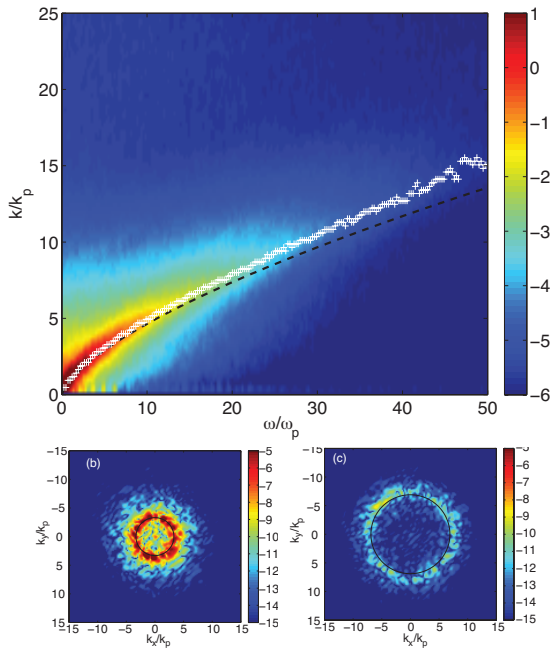}\\
\caption{(a): $S_{\eta}(\omega/\omega_p,k/k_p)$. ($--$): Linear dispersion relation $\omega^2=\frac{\gamma}{\rho}k^3$. (white $+$): spectrum maxima. (b,c): $S_{\eta}(k_x/k_p,k_y/k_p)$ at a fixed $\omega^*=\omega/\omega_p=6$ (b) and $\omega^*=\omega/\omega_p=18$ (c). Circles in solid line indicate $k^*(\omega^*)$ given by the dispersion relation. The wave field is found isotropic. Colors are $S_{\eta}$ log-scaled.}
\label{simu2}
\end{center}
\end{figure}

The space-time wave height spectrum $S_{\eta}(\omega/\omega_p,k/k_p)$ is shown in Fig. \ref{simu2} (a). Energy is found to be localized in the Fourier space around the linear dispersion relation of capillary waves. The local maxima of the spectrum for each frequency (crosses) fall relatively well on the theoretical dispersion relation curve. A slight mismatch between theoretical and numerical values occurs at high frequencies which is attributed to the numerical dispersion (see the supplementary materials \cite{SMDeike} for details) that is linked to the lack of resolution for the highest frequencies. While forcing is only applied at low $k$, the energy spectrum spreads over a large range of wave numbers showing that nonlinear transfers among the different scales have taken place. Moreover, the dispersion relation broadens in frequency as the wave number increases, first due to non-linear interaction between the waves and then at high frequencies ($\omega/\omega_p>20$) due to dissipative effects, as discussed later on. Note also that for simulations with lower values of $\sigma_{\eta}$ the space-time spectrum non-linear broadening is reduced (not shown).  Figures \ref{simu2} (b) and (c) depict the spatial spectrum $S_{\eta}(k_x/k_p,k_y/k_p)$ at two different frequencies $\omega^*$. An isotropic wave field is observed where the energy is localized around a circle of radius $k^*=\sqrt{k_x^{*2}+k_y^{*2}}$ well predicted by the linear dispersion relation $\omega^*(k^*)$. Again a significant broadening on the spectrum is observed for high wave numbers. These observations still holds regardless of the frequency and the forcing amplitude. Thus, numerical computations capture well capillary waves that propagate at various scales in an isotropic wave field.

\begin{figure}
\begin{center}
\includegraphics[scale=0.45]{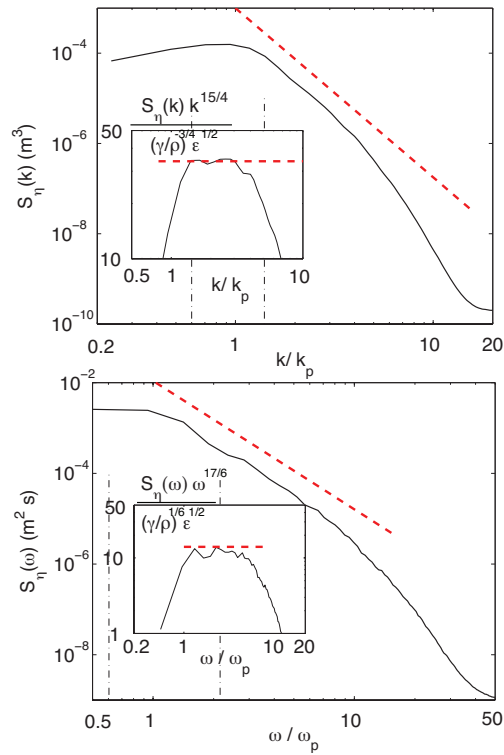}
\caption{(a): $S_{\eta}(k/k_p)$. (b): $S_{\eta}(\omega/\omega_p)$. Theoretical KZ spectrum ($--$) is respectively $S_{\eta}\sim k^{-15/4}$ and $S_{\eta}\sim \omega^{-17/6}$. Inset: compensated spectrum, respectively $S_{\eta}(k)(k)^{15/4}/(\epsilon^{1/2}\left(\gamma/\rho\right)^{-3/4})$ and $S_{\eta}(\omega)\omega^{17/6}/(\epsilon^{1/2}\left(\gamma/\rho\right)^{1/6})$. Vertical dot-dashed line indicates the forcing scale range. }
\label{simu3}
\end{center}
\end{figure}

Figure \ref{simu3}a depicts the spatial spectrum $S_{\eta}(k/k_p)$, obtained by integrating over all frequencies the space-time spectrum depicted in Fig. \ref{simu2}. It exhibits a power-law regime within an inertial range $1\lesssim k/k_p\lesssim 6$ from the forcing scale to a dissipative cut-off length, beyond which the spectrum departs from the power law. The frequency spectrum (obtained by integrating over $k$) also exhibits a power-law in the range $1\lesssim \omega/\omega_p\lesssim 8$ (see Fig. \ref{simu3}b). The inertial range spreads over roughly one decade in frequency (the range is smaller in terms of wave numbers). The observed power laws within the inertial range are found to be in good agreement with wave turbulence theory scalings: $S_{\eta}(k)\sim k^{-15/4}$, $S_{\eta}(\omega)\sim \omega^{-17/6}$ \cite{Zakharov67}. The compensated spectrum is shown in the inset in both cases (Fig. \ref{simu3}a and b). The flat spectrum observed within the inertial range confirms the good agreement between DNS 
and wave turbulence theory.  This limited inertial range is a consequence of the finite resolution of the interface, therefore it may be enlarged at expenses of larger computational resources (see \cite{SMDeike} for details).

It is now possible to estimate numerically the KZ constant from Eq. (\ref{eqKZ}) and the compensated spectrum: $C^{KZ}_{k}=S_{\eta}(k)k^{15/4}/[\epsilon^{1/2}\left(\gamma/\rho\right)^{-3/4}]$ (inset Fig. \ref{simu3}a), or $C^{KZ}_{\omega}=S_{\eta}(\omega)\omega^{17/6}/[\epsilon^{1/2}\left(\gamma/\rho\right)^{1/6}]$ (inset Fig. \ref{simu3}b). To this end, we evaluate the mean energy flux $\epsilon$ using the measure of the dissipated power $D=2\rho\nu\int_{\Omega} S_{ij}S_{ij}d\Omega$, with $S_{ij}$ the deformation tensor $S_{ij}=\frac{1}{2}\sum_{ij}(\partial v_i)/(\partial x_j)$, $i=\{x,y,z\}$ \cite{Pope}. The mean energy flux is then defined by $\epsilon\equiv \langle D\rangle /(A\rho)$, where $A$ is the surface area and $\langle.\rangle$ designed an average over time. A comparable value of the flux is obtained using the dissipation spectrum as in \cite{DeikePre2013}, $\epsilon_d=\int d\omega dk \frac{\gamma}{\rho} k^2 S_{\eta}(k,\omega) /\tau^{emp}$ where $\tau^{emp}= \tau_d$ for $k/k_p<6$ and $\tau^{emp}=\tau_d^{num}$ for $k/k_p>6$, so that $\tau^{emp}$ includes $\tau_d=1/(2\nu k^2)$ \cite{Lamb} the viscous linear dissipation, and $\tau_d^{num}$ the total (numerical and physical) dissipation time valid only at small scales $(k/k_p>6)$ \cite{SMDeike}. We obtain from the DNS the following estimation of the KZ constant, $C^{KZ}_{\omega}= 16$ and $C^{KZ}_{k}=34$. Thus $C^{KZ}_{k}/C^{KZ}_{\omega}\approx 2$, while the theoretical ratio given by the 
relation $S_{\eta}(k)dk=S_{\eta}(\omega)d\omega$ is $3/2$. The KZ constant $C_{nk}$ can be also defined from the wave action spectrum equation, and $C^{KZ}_{k} = 2\pi C_{nk}$ \cite{Pushkarev2000}. Using the $C^{KZ}_k$ value from our simulations and the latter equation leads to $C_{nk} = 5\pm 1$, the uncertainty coming from using either the value of $C^{KZ}_k$ or $C^{KZ}_{\omega}$. Our estimation of $C_{nk}$ is of the same order of magnitude as $9.85$ the theoretical value found in \cite{Pushkarev2000}. The difference between these two values can be due to the short inertial range induced by the numerical dissipation. Note that in the previous Hamiltonian simulations, the KZ constant was found two times smaller than ours \cite{Pushkarev2000}. This difference is probably related to the limited length of the inertial range and the numerical dissipation at small scales, our numerical methods allowing for a better resolution of these scales.

\begin{figure}
\begin{center}
\includegraphics[scale=0.45]{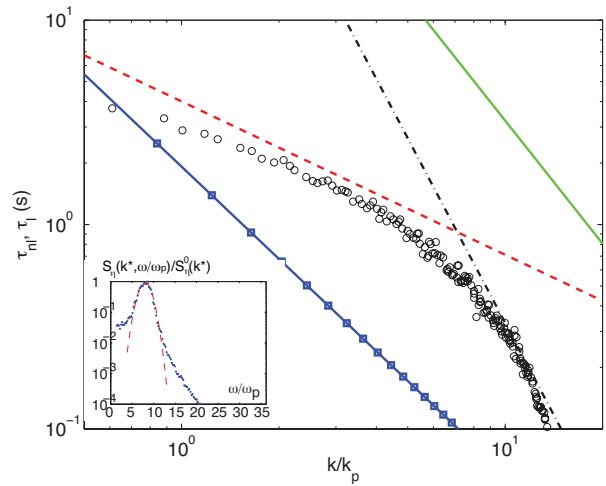}
\caption{Time scale evolutions as a function of $k/k_p$. ($\circ$): $\tau_{nl}(k)=1/\Delta\omega$; $(\blacksquare -)$: $\tau_l=1/\omega$; $(-)$: $\tau_d=1/(2\nu k^2)$ viscous dissipation and $(-\cdot -)$: $\tau_d^{num}$ numerical dissipation function ($1/\tau_d^{num}=\mathcal{A}\frac{k}{k_{max}}/\tau_d$, with $\mathcal{A}\approx 100$ an empirical non dimension parameter \cite{SMDeike}). Theoretical prediction $\tau_{nl}\sim k^{-3/4}$ ($--$). Inset: $S_{\eta}(k^*,\omega)$ normalized by its maximum $S_{\eta}^0(k^*)$ as a function of $\omega/\omega_p$, for $k^*/k_p=7.6$. $(--)$: gaussian fit, $\exp{[-(\omega -\omega_c)^2/\Delta\omega^2]}$ with $\Delta \omega=2\pi\ 0.15$ and $\omega_c=2\pi\ 0.66$ rad$^{-1}$.}
\label{simutnl}
\end{center}
\end{figure} 

One of weak turbulence key hypothesis regards the time scale separation. The linear wave propagation time is indeed supposed to be much smaller than the time scale of the nonlinear energy exchanges. This latter must also be small compared to the dissipative time. We will now evaluate the various time scales involved in the problem. As already discussed, $S_{\eta}(k,\omega)$ broadens around the linear dispersion relation curve. The interaction nonlinear time scale $\tau_{nl}(k)$ is linked to this broadening \cite{NazarenkoBook,Miquel2012,Deike2013}. This time is defined by $\tau_{nl}(k)=1/\Delta\omega(k)$, where $\Delta \omega(k)$ is the inverse of the spectrum width at a given wave number $k$. As shown in the inset of Fig. \ref{simutnl}, $\Delta \omega(k)$ is extracted from $S_{\eta}(k,\omega)$ using the rms value of a Gaussian fit with respect to $\omega$ at a given $k^*$. Then, iterating this protocol to all $k^*$ allow us to determine $\tau_{nl}(k)$.  Figure \ref{simutnl} shows that $\tau_{nl}(k)$ is 
found to be close to the theoretical scaling of capillary wave turbulence $\tau_{nl}\sim k^{-3/4}$  \cite{Zakharov67} within the inertial range $1<k/k_p<6$, and then strongly decreases for larger $k$. The linear propagation time $\tau_l=1/\omega$ is also shown on Fig. \ref{simutnl} where we see a clear time scale separation $\tau_l\ll \tau_{nl}$ within the inertial range. Close to the forcing scales, both times are of the same order of magnitude. The dissipative time scales are also shown, the solid line indicates the classic viscous linear dissipation $\tau_d=1/(2\nu k^2)$ \cite{Lamb}, while the dot-dashed line displays the extrapolated empirical (physical and numerical) dissipation $\tau_d^{num}$, determined by measuring the decay rate of a 2D freely decaying capillary wave for various spatial resolution  \cite{SMDeike}. Thus when the nonlinear time scale becomes of the same order as the total dissipation time, the cascade progressively ends and dissipation is responsible for the broadening of the spectrum, as already observed in Fig. \ref{simu2}. We note that the numerical dissipation does not affect the capillary wave turbulence cascade within the inertial range, while the physical dissipation would become dominant for very high resolution far beyond current computational resources. As expected by the weak turbulence theory, the power law spectrum observed is shown to fall within the range defined by the double inequality $\tau_l\ll \tau_{nl}\ll \tau_d^{num}$.

In conclusion, this work presents DNS of capillary wave turbulence where the two-phase 3D Navier Stokes equations have been solved. The wave height spectrum is found to exhibit frequency and wave number power-laws in good agreement with weak turbulence theory. We also observe a clear time scale separation between linear and nonlinear times. These numerical results confirm the validity of weak turbulence approach to quantify out-of-equilibrium wave statistics. It also opens new perspectives in order to better understand wave turbulence systems and the influence of the air and water flows. For instance, further numerical results should shed new insight on the importance of dissipation at all scales as recently reported experimentally in capillary \cite{DeikePre2013} and flexural \cite{Humbert2013,Miquel2013} wave turbulence. Moreover the inclusion of gravity in the present simulations would allow to investigate the role of the gravity-capillary transition, strongly nonlinear structures (wave breaking, soliton, ...), as well as gravity wave turbulence.

\begin{acknowledgments}
We thank S. Popinet and C. Josserand for discussions, as well as the Institut Jean le Rond d'Alembert for their computational facilities. This work has been partially supported by ANR Turbulon 12-BS04-0005.
\end{acknowledgments}
\bibliographystyle{apsrev4-1}
\bibliography{wtdns.bib}

\end{document}